\documentstyle[12pt,subeqn,fullpage,epsf]{article}
 
\def\lsim{\:\raisebox{-0.5ex}{$\stackrel{\textstyle<}{\sim}$}\:}
 
\def\be{\begin{equation}}
\def\ee{\end{equation}}
\def\beq{\begin{eqnarray}}
\def\eeq{\end{eqnarray}}
\def\L{\cal L}
\begin{document}
\begin{center}
{\Large{\bf Seeking supersymmetry at LEP}} \\[3cm]
{\large{\bf Probir Roy}} \\[.5cm]
Tata Institute of Fundamental Research \\
Homi Bhabha Road, Bombay 400 005, India \\[1cm]
{\it Invited talk given at the Workshop on High Energy Physics
Phenomenology 4}\footnote{S.N. Bose National Centre for Basic
Sciences, Calcutta, India, Jan. 2-16, 1996}, {\it to appear in its
Proceedings} \\[3cm] 
\end{center}
\begin{itemize}
\item {\bf Scope and motivation}
\item {\bf Lightning review of CMSSM}
\item {\bf LEP 1 bounds and LEP 1.5 results}
\item {\bf LEP 2 futurology}
\item {\bf Bottomline}
\end{itemize}
\newpage

\begin{itemize}
\item {\bf Scope and motivation}
\end{itemize}

This minireview has been commissioned by the organizers of the
workshop. 
Under the general title of LEP, I aim to cover three segments of the
experimental programme in that accelerator.  First, I consider the
analyzed sample of $Z$-events out of the 20 million collected at LEP
1.  The lack of observation of any direct supersymmetry signal in
these has yielded bounds on supersymmetric parameters.  Second, there 
is about $5~({\rm pb})^{-1}$ of data which have been collected
during the intermediate energy run (LEP 1.5) at 130, 136 and 140 GeV in
the $e^+e^-$ centre of mass.  A preliminary examination of these data
is revealing for supersymmetry search strategies.  Finally, I briefly
indulge in 
some futurology vis-a-vis the coming runs in LEP 2 which will be in
the energy regime 161-200 GeV.

Our supersymmetry reference point is the supergravity-constrained
\cite{one} minimal supersymmetric standard model CMSSM --
distinguished by the possession of the smallest number of
parameters among all contending SUSY scenarios.  Sometimes, a less
strong version of this model is considered \cite{two} in which one
partially relaxes the supergravity-based unification assumption of a
universal 
supersymmetry-breaking scalar mass at the scale by taking separate
scalar mass 
values for the Higgs and the sfermion sectors; this version will be
differentiated by using the prefix ``partially constrained'': PCMSSM.
Once in a while, we shall also mention the supergravity-constrained
next-to-minimal supersymmetric standard model CNMSSM \cite{three},
possessing an additional gauge-singlet superfield $N$ whose scalar VEV
$\langle n\rangle$ leads to the Higgsino mass parameter $\mu$.

The motivation for considering sub-TeV to TeV scale supersymmetry in
particle 
physics is well-known.  It naturally ensures the stability of the weak
scale against radiative perturbations induced from unknown high scale
physics.  It has a pronounced decoupling feature in that at low
energies it reduces to the Standard Model with a somewhat light
physical Higgs scalar 
when all sparticles are made heavier \cite{four} than about 200 GeV.
As a result, all the presently successful features of the Standard
Model are easily retained (in the sense of being compatible within
errors) in the supersymmetric extension.  A particularly notable item
among the latter is quantum consistency at the 1-loop level which
predicts \cite{five} the top quark mass $m_t$ to be $169 +
7^{+4}_{-3}$ GeV to be compared with the Tevatron-measured [summer,
1996] value of $175 \pm 9$ GeV from CDF and $170 \pm 18$ GeV from
${\rm D}0\!\!\!/$.  In addition, though, supersymmetry
provides the standard model with a bridge from electroweak energies to
very high scale physics, e.g. $M_X \sim 3 \times 10^{16}$ GeV
\cite{six} of supersymmetric grand unification or $M_S \sim 5 \times
10^{17}$ GeV of superstring unification \cite{seven} or the Planck
scale $M_P \sim 1.2 \times 10^{19}$ GeV of quantum gravity.

Apart from the above virtues, the CMSSM is characterized by a powerful
simplicity as well as phenomenological tractability and definite
testability.  As many as 31 parameters, which appear \cite{eight} in
the straightforward minimal extension of the low-energy Standard Model
to its supersymmetric version (MSSM) with soft supersymmetry breaking
(SSB) 
\be
\L^{SM} \rightarrow \L^{SSM} + \L^{SSB},
\ee
are reduced in the supergravity-constrained MSSM to 4 plus a sign --
namely that of $\mu$.  (In PCMSSM one has 5 parameters including the
magnitude of $\mu$).  Moreover,
there are well-formulated tests, the violation of any of which will
kill the model.  As an illustration, one may mention the possible
observation of a slepton heaver than the corresponding squark (in the
first two generations) which will destroy CMSSM.  Another killer would
be the 
demonstration of the absence of any neutral Higgs particle with a mass
\cite{ten} below about 130 GeV.  The latter, in fact, would exclude
MSSM itself.

Let us focus on the crucial assumptions that go into the CMSSM chosen
here and its salient features.  To start with, there is a spectrum 
comprising particles and their partner sparticles.  The particles are
those of the 
Standard Model except that, instead of one, there are five physical
Higgs scalars (two charged 
$H^\pm$, two CP-even neutrals -- the lighter $h$ and the heavier $H$
-- and one CP-odd neutral $A$) coming from two Higgs doublets.  (The
ratio of the VEV of the neutral Higgs field which couples with
up-type fermions to that of the one which does so with down-type
fermions -- is called $\tan\beta$).  The 
generic sparticle is expected to be heavier than the corresponding
particle,  
though the order could be reversed for the right-chiral stop and/or
the chargino, neutralino.  All coupling and mass parameters evolve with
the energy scale -- described by 26 independent renormalization group
equations.  Next, postulated boundary conditions are imposed on some
of these parameters at the unification scale $M_X \sim 2 \times
10^{16}$ GeV.  Specifically, all supersymmetry-breaking scalar
(gaugino) masses are taken to be universal and equal to one mass $m_0$
($M_{1/2}$).  Squared masses of the Higgs at the unification scale of
course, have an additional supersymmetric contribution, namely
$\mu^2$,  since the higgsino mass is also a supersymmetric
contribution to the Higgs mass.
Again, all supersymmetry breaking trilinear scalar couplings,
$A_{ijk}$ ($i,j,k$ generation indices) are taken to be equal $(= A_0)$
at $M_X$.  Here $m_0$ and $M_{1/2}$ 
are supposedly of the order of the gravitino mass $m_{3/2} \sim
\wedge^2_S M^{-1}_{P\ell}$, with $\wedge_S$ being some dynamical
supersymmetry-breaking scale in the hidden sector $\sim 10^{11} -
10^{12}$ GeV.  Now $m_0, A_0, m_{1/2}$ and $\tan\beta$ can be chosen
to be the four parameters; or, $m_A$ could be traded for one of the
first two.    
The Higgs squared masses evolve rapidly in a decreasing
manner as one lowers the energy scale.  The starting values of the
parameters 
at the unification scale ensure that one such squared mass of a
neutral Higgs flips its sign, while evolving, thus leading to
radiative electroweak symmetry 
breakdown and a weak scale $M_W \lsim O(m_{3/2})$.  Though CMSSM still
has a large allowed region in its space of parameters, we will -- from
time to time -- talk about two extreme scenarios just to emphasize their
difference: 

\begin{enumerate}
\item[{\#}] Decoupling SUSY ---- all sparticles heavier than 200
GeV.  
\item[{\#}] SUSY round the corner ---- a stop, a chargino $+$ a
neutralino, each below 100 GeV.
\end{enumerate}

\noindent In the less ambitious PCSSM, universality at $M_X$ is
imposed only on squark and slepton (and not on the Higgs) masses --
thus keeping an open mind on the radiative origin of the Higgs
mechanism.  {\it Caveat emptor}: Some of the highly restrictive
assumptions underlying CMSSM may be plain wrong!  
\bigskip

\begin{itemize}
\item {\bf Lightning review of CMSSM}
\end{itemize}

The superfield content of the model in the matter sector, written in a
transparent notation ($i = 1,2,3$ is a generation index), is:
\[
Q^i_L = \left(\matrix{U^i_L \cr D^i_L}\right), ~L^i_L =
\left(\matrix{N^i_l \cr E_L}\right), ~Q^i_R, L^i_R, H_1 =
\left(\matrix{H^0_1 \cr H^-_1}\right), ~H_2 = \left(\matrix{H^+_2 \cr
H^0_2}\right).  
\]
The corresponding superpotential (with $R$-parity assumed conserved)
is:
\be
W = \lambda^{ij}_U Q^i_L\!\!\cdot\!\!H_2 U^j_R + \lambda^{ij}_D
Q^i_L\!\!\cdot\!\!H_1
D^j_R + \lambda^{ij}_E L^i_L\!\!\cdot\!\!H_1 E^j_R + \mu
H_1\!\!\cdot\!\!H_2, 
\ee
with $\lambda$'s as Yukawa couplings.  

The scalar potential can be derived from (1).  Writing $\phi_j$ for a
generic scalar field and incorporating the soft supersymmetry breaking
terms, we have
\beq
V &=& \sum_j \left|{\partial W \over \partial \phi_j}\right|^2 +
D{\rm -terms} + \sum_{i,j} m^2_{ij} \phi_i \phi_j \nonumber
\\[2mm] & & + \left\{A_U \lambda_U \tilde q_L\!\!\cdot\!\!h_2 \tilde u_R +
A_D \lambda_D \tilde q_L\!\!\cdot\!\!h_1 \tilde d_R + A_E \lambda_E \tilde
e_L\!\!\cdot\!\!h_1 \tilde e_R + B\mu h_1\!\!\cdot\!\!h_2 + {\rm
H.C.}\right\}.  
\eeq
In (3) the third RHS term includes $m^2_1 h^+_1 h_1 + m^2_2 h^+_2 h_2$
with a vanishing $m_{12}$ where  
$h_{1,2}$ refers to the scalar component of the superfield
$H_{1,2}$.  Also, $v_{1,2} = \langle h^0_{1,2} \rangle$ and $\tan\beta
= v_2/v_1$. The physical fields can be expressed in terms of
the superfield components given above.  For instance, the field for
the lightest neutral scalar is $h = \sqrt{2}({\rm Re}~h^0_2 - v_2)
\cos\alpha - \sqrt{2}({\rm Re}~h^0 - v_1) \sin\alpha$, where $\alpha$
is an 
angle which enters via mixing.  The orthogonal heavier combination is
$H = \sqrt{2} ({\rm 
Re}~h^0_2 - v_2) \sin\alpha + \sqrt{2}({\rm Re}~h^0_1 - v_1)\cos\alpha$ 
while $A$ equals $\sqrt{2} ({\rm Im}~h^0_2 \cos\beta - {\rm Im}~h^0_1
\sin\beta)$.  The partners of the CKM
matrices in the scalar sector are assumed to
possess safety properties which suppress dangerous flavor-changing
neutral current processes that could emerge from (1).

At the tree level itself one has several mass relations.
\subequations
\be
m^2_\pm = m^2_A + M^2_W,
\ee
\be
m^2_h \leq M^2_Z \leq m^2_H,
\ee
\be
{m_h \over |\cos 2\beta|} ~<~ m_A ~<~ m_H,
\ee
\be
\mu^2 = (\cos2\beta)^{-1} (m^2_2 \sin^2\beta - m^2_1 \cos^2\beta) -
{1\over2} M^2_Z,
\ee
\be
2B\mu = (m^2_1 - m^2_2) \tan2\beta + M^2_Z \sin2\beta,
\ee
\be
m^2_A = m^2_1 + m^2_2 + 2\mu^2.
\ee
\endsubequations

On including 1-loop quantum corrections in the leading log
approximation, the upper bound on the 
squared mass of $h$ reads ($\tilde t_{1,2}$ are the two physical
squarks, assumed to weigh more than the top) \cite{nine}:
\be
M^2_h ~<~ M^2_Z \cos^2 2\beta + {3\alpha_{EM} \over 2\pi
\sin^2\theta_W} {m^4_t \over M^2_W} \ell n {m_{\tilde t_1} m_{\tilde
t_2} \over m^2_t} \simeq (130~{\rm GeV})^2.
\ee
The boundary conditions at $M_X$ imply
\be
m^2_1 (M_X) = m^2_2 (M_X) = m^2_0.
\ee
Turning to gaugino masses $M_i$ ($i =$ nonabelian gauge group index) and
considering 1-loop RGE effects, one can write -- with $\alpha_u$ as
the unified fine structure coupling --
\be
M_i (Q) = M_{1/2} \alpha_i (Q) \alpha^{-1}_u (M_X).
\ee
For the $U(1)_Y$ case, with the standard definition of $Y$, there is
an extra factor of $5/3$ in the RHS.  
It turns out that $M_1$ $(M_Z) \simeq 0.41~M_{1/2}$ and $M_2$ $(M_Z)
\simeq 0.84~M_{1/2}$ with a mild $Q$-dependence in $M_{1,2}$.
However, the situation is quite different for $M_3$.  The physical
on-shell gluino mass $m_{\tilde g}$ is given by \cite{ten}
\beq
m_{\tilde g} = M_3 (Q) \Bigg[1 &+& {\alpha_S (Q) \over 4\pi} \Big\{15 -
18 \ell n {M_3 (Q) \over Q} \nonumber \\[2mm]
&+& \sum_q \int^1_0 dx ~x~\ell n {xm^2_q +
(1-x)m^2_{\tilde q} - x(1-x)M^2_3 \over Q^2}\Big\}\Bigg]
\eeq
and is independent of $Q$.  For $M_3 \simeq 0.1$ TeV and $m_{\tilde q}
\simeq 1$ TeV, the difference between $m_{\tilde g}$ and $M_3 ~(M_3)$
can be as much as 30\%.

The spectrum of the remaining sparticles can be parametrized, after
accounting for renormalization group evolution, as
follows \cite{eleven}:
\subequations
\be
m^2_{\tilde e_R} = m^2_0 + 0.15 M^2_{1/2} - \sin^2 \theta_W D;
\ee
\be
m^2_{\tilde e_L} = m^2_0 + 0.52 M^2_{1/2} - \left({1\over2} -
\sin^2\theta_W\right) D;
\ee
\be
m^2_{\tilde\nu} = m^2_0 + 0.52 M^2_{i/2} + {1\over2} D;
\ee
\be
m^2_{\tilde q\uparrow R} = m^2_0 + (0.07 + C_{\tilde g})M^2_{1/2} +
{2\over 3} \sin^2\theta_W D;
\ee
\be
m^2_{\tilde q\downarrow R} = m^2_0 + (0.02 + C_{\tilde g})M^2_{1/2}
- {1\over3} \sin^2\theta_W D;
\ee
\be
m^2_{\tilde q\uparrow L} = m^2_0 + (0.47 + C_{\tilde g}) M^2_{1/2} +
\left({1\over2} - {2\over3} \sin^2 \theta_W\right)D;
\ee
\be
m^2_{\tilde q\downarrow L} = m^2_0 + (0.47 + C_{\tilde g})M^2_{1/2} -
\left({1\over2} -{2\over3} \sin^2\theta_W\right)D.
\ee
\endsubequations
Here $C_{\tilde g} = {8\over9} \left[\alpha^2_S (m_{\tilde
q})/\alpha^2_S (M_X) - 1\right]$ and $D = M^2_Z \cos^2\beta$ while we
have $\ell = e,\mu$, $q\!\!\uparrow = u,c$ and $q\!\!\downarrow =
d,s,b$.  For 
stops and staus, considerable left-right mixing is anticipated.  The
corresponding mass-squared matrices are given by
\subequations
\be
m^2_{\tilde t} = \left(\matrix{m^2_{\tilde q_{\uparrow L}} + m^2_t
+ 0.35 D & 
-m_t(A_t + \mu \cot \beta) \cr -m_t (A_t + \mu \cot \beta) &
m^2_{\tilde q \uparrow R} + m^2_t + 0.16 D} \right),
\ee
\be
m^2_{\tilde \tau} = \left(\matrix{m^2_{\tilde \ell_L} + m^2_\tau -
0.27 D & 
-m_\tau (A_\tau + \mu \tan\beta) \cr -m_\tau (A_\tau + \mu \tan\beta) &
m^2_{\tilde \ell_R} + m^2_\tau - 0.23 D} \right).
\ee
\endsubequations
A sample scatter plot of the ranges \cite{twelve} of some characteristic
masses in the model -- showing the extent of variation in the
parameter space -- is shown in Fig. 1.  One should also mention 

\newpage
\begin{center}
\begin{figure}[hbt]
\vskip -1in
\begin{center}\leavevmode
\epsffile[50 350 260 420]{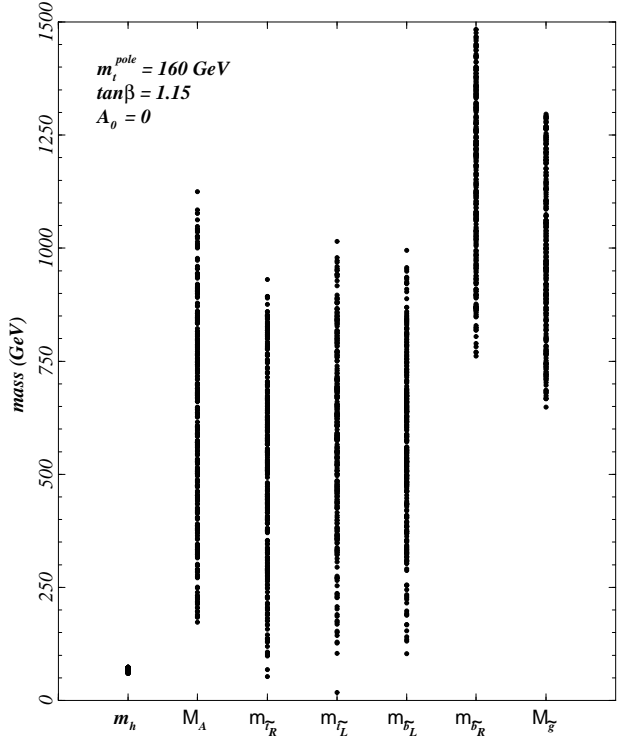}\end{center}%
\vskip 0.75cm
\vskip 9.5cm
\caption[Ranges of some sparticle masses] 
        {Ranges of some sparticle masses} 
\end{figure}
\end{center}

\noindent that five squarks (i.e. all except the stop)
need to be taken as nearly mass-degenerate in order to avoid an
unacceptable FCNC-induced $K^0 - \bar K^0$ mixing.  This could be a
problem in Fig. 1 \cite{twelve} which has a rather large $\tilde b_L -
\tilde b_R$ mass-splitting.  A similar argument
vis-a-vis the FCNC-induced $\mu \rightarrow e\gamma$ decay requires
the near mass-degeneracy of all sleptons except $\tilde \tau$.
\bigskip

\begin{itemize}
\item{\bf LEP 1 bounds and LEP 1.5 results}
\end{itemize}

{\it Higgs constraints}
\smallskip

Let me first reiterate the experimental mass bounds on the neutral Higgs
scalars.  The SM Higgs particle has been searched for in the Bjorken
process $e^+e^- \rightarrow Z \rightarrow h Z^\star \longrightarrow
b\bar b \ell(q) \bar\ell (\bar q)$.  The present SM lower bound of
63.5 GeV gets diluted significantly in case of $h$ of MSSM since $\Gamma^{\rm
MSSM}_{Bj} = \sin^2 (\alpha - \beta) \Gamma^{SM}_{Bj}$ and the
situation has not been helped by the LEP 1.5 data.  The exact lower
bound on $m_h$ in MSSM is presently a matter of controversy.  
The CP-odd scalar $A$ has also been looked for in the process $e^+e^-
\rightarrow Z \rightarrow hA$.  By use of the formula
\be
\Gamma (Z \rightarrow hA) = {1\over2} \cos^2(\alpha-\beta) \Gamma(Z
\rightarrow \nu\bar\nu) \left[\lambda\left(1,m^2_h
M^{-2}_Z,m^2_AM^{-2}_Z\right)\right]^{3/2}, 
\ee
the lack of observation of any $A$ translates to the bound $m_A ~>~
26$ GeV for $1 ~<~ \tan\beta ~<~ 50$, assuming that $h$ allows the
process kinematically.  It is worth pointing out in
this context that no MSSM Higgs production interpretation can be given
to the excess $4j$ events reportedly seen by ALEPH at LEP 1.5 owing to
the alleged lack of $b$'s in the final state.

\newpage

\noindent {\it Direct sparticle mass bounds}
\smallskip

Charginos and charged sfermions are open to pair-production from the
resonant $Z$.  The signals would be hard acollinear jets (and or
leptons) and ${E\!\!\!/}_T$.  The lack of observation of any such
event at LEP 1 implies
that those sparticles are all heavier than $M_Z/2$.  In LEP
1.5, considering the offshell $Z$, the relevant lower bound would be
$\sqrt{s}/2$.  One also needs to take into
account the experimental constraint that any possible partial width of
the $Z$ decaying into visible channels must obey $\Delta \Gamma_Z$
(new and visible) $<~0.13$ MeV.  All these constraints, combined
within PCMSSM and assuming two light neutralinos, exclude certain regions
of the $\mu - \tan\beta$ plane. 
We show these exclusion zones \cite{thirteen} in Fig. 2.
Fig. 2a shows the situation for the decoupling case where all
superparticles, except the lightest chargino, weigh more than 200 GeV;
the boundaries of the dashed region, already excluded by LEP 1 data,
would expand further as shown in case there is no chargino upto 90
GeV.  A similar plot is made for the ``round the corner'' scenario 
in Fig. 2b.

\centerline{}
\vskip -3.0cm
\overfullrule=0pt
\begin{figure}[hbt]
\begin{minipage}[h]{2.2in}
\begin{center}\leavevmode
\epsffile[185 390 432 560]{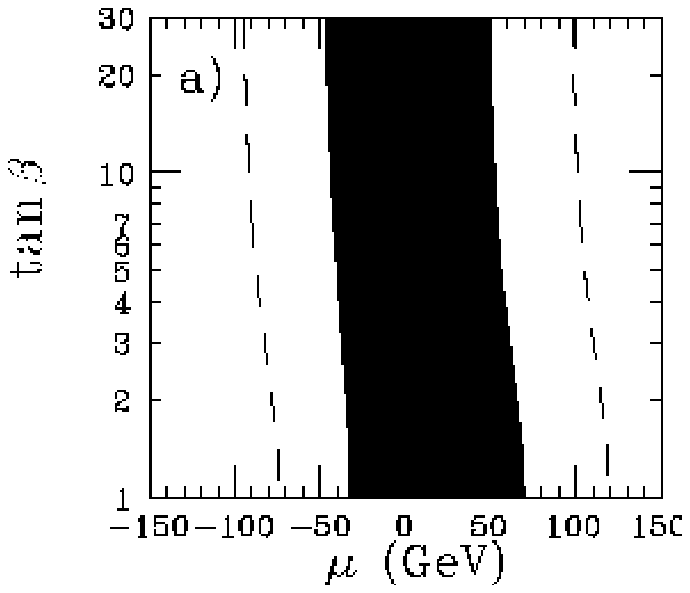}\end{center}%
\end{minipage} \hfill
\begin{minipage}[h]{3.4in}
\begin{center}\leavevmode
\epsffile[170 312.2 432 560]{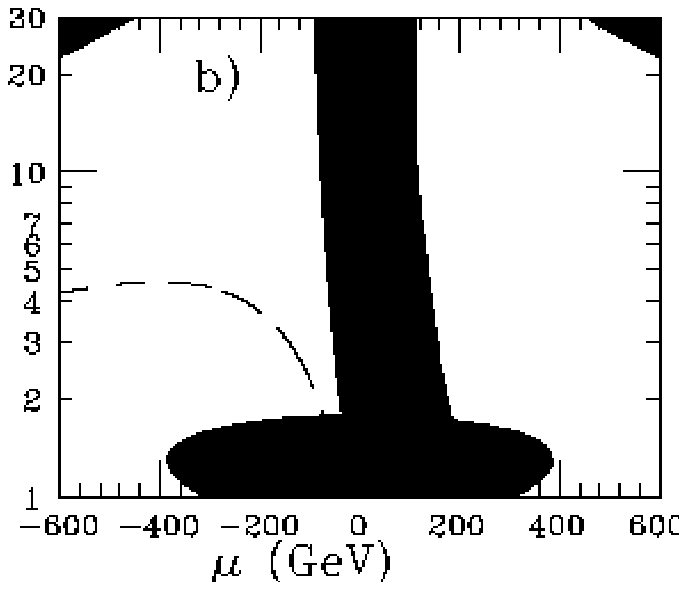}\end{center}%
\end{minipage}
\vskip 0.75cm
\caption[Combined LEP restriction in $\mu$ {\it vs.} $\tan\beta$ plane]
{ PCMSSM exclusion regions in the $\mu - \tan\beta$ plane}
\label{lepcommutb}
\end{figure}
\overfullrule=5pt
\vskip 0.5cm

Let me come next to neutral sparticles, such as sneutrinos and
neutralinos.  One needs to first consider the nonaccelerator constraints
on the former.  Stable sneutrinos in the range 3 GeV to 1 TeV are
practically 
excluded by dark-matter search experiments.  Unstable ones must weigh
more than 41.8 GeV, otherwise they would allow the decay chain $Z
\rightarrow \tilde \nu \tilde\nu^\star$, $\tilde\nu \rightarrow \nu +
{\rm LSP}$ violating the LEP bound $\Delta \Gamma_Z$ (new and
invisible) $<6.7$ MeV.  Turning to the neutralinos
$(\tilde\chi_{1,2,3,4})$, we note that, among the modes $Z \rightarrow
\tilde\chi^0_i \tilde\chi^0_j$, the $\tilde \chi^0_1 \tilde\chi^0_1$
final state is undetectable and that the sum of the branching ratios
of all such detectable 
processes is experimentally constrained to be less than $10^{-5}$.
The simplest produced final state is $\tilde\chi^0_1 \tilde\chi^0_2$
with the subsequent decay $\tilde\chi^0_2 \rightarrow \tilde\chi^0_1 +
{\rm visible}$.  There can, however, be other states \cite{fourteen} such as
$\tilde\chi^0_3 \tilde\chi^0_4$, each of which can decay into
$\tilde\chi^0_1 + Z$ or $h$.  Analyzing the LEP data, one can
conservatively quote 
$M_{\tilde\chi^0_1} > 23$ GeV in the CMSSM and PMSSM, but the lower
bound reduces to $0$ in CNMSSM.  For $\tilde\chi^0_2$, a stronger statement,
namely $M_{\tilde\chi^0_2} > 56$ GeV is possible \cite{fourteen}.
\bigskip

\newpage

\noindent {\it Indirect constraints}
\smallskip

These are obtained by inputting $\Gamma^Z_{tot.}$, $\Gamma^Z_{ee}$,
$\sin^2 \theta^e_{EFF}$, $A^b_{FB}$, $A^c_{FB}$, $\Delta r$, $R_\ell$,
$R_b$ and $\Delta r \equiv 1 - \pi\alpha_{EM}(\sqrt{2} G_F)^{-1} M^{-2}_W
(1 - M^2_W M^{-2}_Z)^{-2}$ from experiment.  The 
data (at 90\% c.l.) allow a band, as plotted against an MSSM mass
parameter, say the mass of the CP-odd Higgs $m_A$ (Fig. 3).
Three typical PMSSM fits are plotted \cite{fifteen}.  The
bold line corresponds to $\tan\beta = 70$, the dashed one to
$\tan\beta = 20$, the dashed-dotted to $\tan\beta = 8$, the
long-dotted to $\tan\beta = 1.5$ and the dotted one to $\tan\beta =
0.7$.  The average 
squark and slepton masses have been fixed at 900 GeV and 500 GeV
respectively while the gluino mass has been kept at 800 GeV.  The
$\mu$-parameter and $M_{1/2}$ have been kept at $-100$ GeV and 300 GeV
respectively and no $L$-$R$ mixing has been assumed.  Evidently, all
quantities except $R_b$ and $R_c$ can be fit naturally and easily;
however, in these MSSM does not necessarily do better than SM.  

Focusing on $R_b$ and $R_c$, the experimental numbers quoted in
conferences during the summer of '95 are:
\[
R_b = 0.2219 \pm 0.0017 ~({\rm c.f. ~SM~value~0.2157, ~i.e.}~
+ 3.7 \sigma ~{\rm deviation}),
\]
\[
R_c = 0.1543 \pm 0.0074 ~({\rm c.f.~SM~value~0.171, ~i.e.}~
- 2.5 \sigma ~{\rm deviation}).
\]
Furthermore, assuming that $R_c$ is given by the SM value, one finds
the experimentally determined $R_b$ to be $0.2206 \pm 0.0016$.  No set
of parameters within MSSM can fit $R_c$ consistently with other
measured quantities.  Ignoring $R_c$, several attempts have been made
to fit $R_b$ through 1-loop enhancements.  Broadly, there are two
alternative scenarios:  (1) the low $\tan\beta$, low mass $(<90~{\rm
GeV})$ chargino and stop option or (2) the large $\tan\beta$, low mass
CP-odd scalar $A$ option.  

For (1) \cite{sixteen}, one makes use of the one loop vertex
correction with the supersymmetric analogue of the top-top-charged
Higgs triangle, namely the stop-stop-chargino triangle.  The
couplings are proportional to $\cot^2\beta$ and the low-mass stop and
chargino enhance the propagator contributions to the second triangle.
Thus $R_b$ can be enhanced.  However, several cautionary remarks are
in order.  First, $\Gamma(b \rightarrow s\gamma)$ has to be kept under
control -- a task not easily done.  Second, with such light stops, the
decay $t \rightarrow \tilde t + {\rm LSP}$ may become too large,
creating problems with the CDF/D$0\!\!\!/$ data.  Finally, the
lack of 
observation of any chargino at LEP 1.5 has poured cold water on
this option.  Claims have also been made that the $\alpha_S(M_Z)$
problem (namely, the discrepancy between its values determined from
LEP and from lower energy measurements) 
is alleviated in this scenario, the SM $Z$-lineshape determination of
$\alpha_S = 0.126 \pm 0.006$ being pulled down to $0.112 \pm 0.005$.

Turning to (2), the saviour \cite{seventeen} is the triangle with
$b$, $b$ and $A$ internal lines.  This will enhance the $A$-induced
1-loop contribution to 
$\Gamma(b \rightarrow c\tau\bar\nu_\tau)$ which needs to be
controlled.  Moreover, one will have the decay $Z \rightarrow b\bar bA$
with the $A$ decaying further into $b\bar b$.  This has not been seen.
Using these two constraints, Wells and Kane \cite{seventeen} claimed
to have excluded this option.  However, there is the question of
experimental efficiency in detecting a four $b$ final state in
$Z$-decay.  Taking that into account, the option is still viable
\cite{eighteen}. 

\newpage
\hrule width 0pt

\centerline{}
\vskip -11.0cm
\hrule width 0pt
\vspace{3.7in}
\begin{center}
\centerline{}
\overfullrule=0pt
\begin{figure}[hbt]
\begin{picture}(125,100)(0,0)
\put(210.0,64.0){\rule{.5in}{.25in}}
\end{picture}
\begin{center}\leavevmode
\epsffile[60 390 488 560]{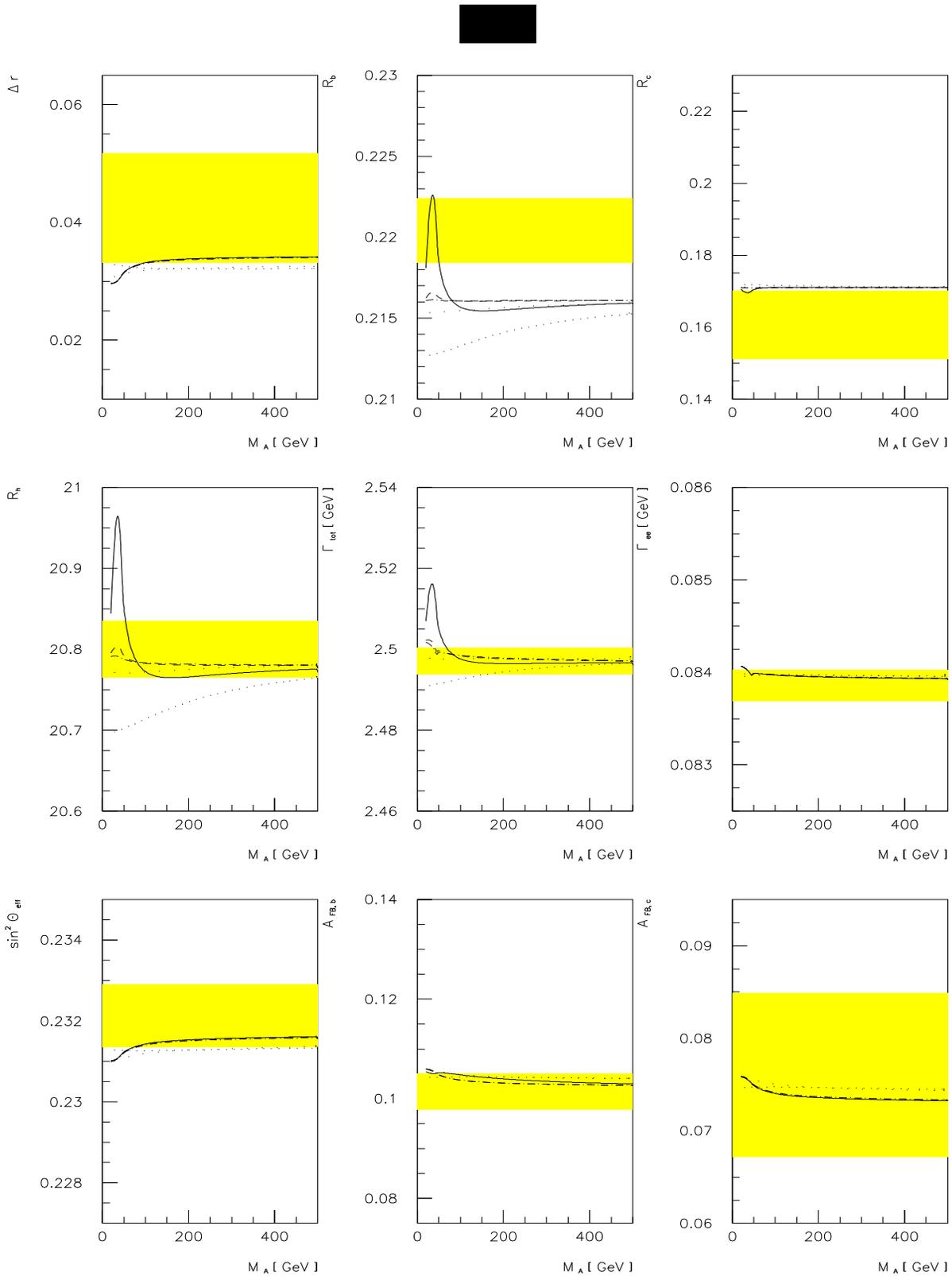}\end{center}%
\vskip 0.75cm
\vskip 11.5cm
\caption[ Electroweak observables plotted against $m_A$ ]
        { Electroweak observables plotted against $m_A$ }
\label{EWobs}
\vskip -10.0cm
\end{figure}
\overfullrule=5pt
\end{center}

\newpage


\begin{itemize}
\item {\bf LEP 2 futurology}
\end{itemize}
\smallskip

The available CM energy of this machine will be in the $161 - 200$ GeV
range.  The dominant SM process will be $WW$ production for which the
cross section has been calculated to be nearly 18 pb, in comparison
with a chargino pair production cross section (for a chargino that is
nearly mass-degenerate with the $W$) in the vicinity of 5 pb.  This
reaction will easily yield three types of final state configurations
which are usually associated with sparticles : (1) acollinear lepton-pair
with $E\!\!\!/_T$, (2) acollinear jets with $E\!\!\!/_T$ and (3)
acollinear jets $+$ 
lepton with $E\!\!\!/_T$.  Thus, $W$-pair production will provide a
severe background to supersymmetry search at LEP 2.  Nevertheless,
there are three redeeming features \cite{fourteen}.  First, the rates
of the actual 
modes that are to be detected need to be calculated by folding
branching ratios with the pair production cross section; after that is
done, 
the nonsupersymmetric channels do not fare particularly better over
their supersymmetric rivals.  Second the geometries of the two types
of events differ significantly; those generated from $W$-pairs would
tend to be anisotropic with a forward peak in contrast
with the 
supersymmetric ones which are more isotropic.  Finally, the kinematics are
very different, the $W$ will largely have a 2-body decay with
$E\!\!\!/_T$ being associated with a massless particle whereas the
chargino will have a 3-body decay with the $E\!\!\!/_T$ coming from 
massive LSP;  intelligently
designed cuts will help discriminate between signal and background.

Another helpful procedure \cite{fourteen} will be a
stage-by-stage increase in energy.  This will keep the requisite
number of open channels in cheek and will faciliate the search
process.  Thus, as $\sqrt{s}$ is increased, one expects first the
pair-production of charginos \cite{nineteen}
$\tilde\chi^\pm$ and neutralinos \cite{twenty}
$\tilde\chi^0$.  These cross sections are expected to be in the
several picobaran range.  At a higher energy, one may expect
right-chiral slepton pair production as the next process.  Here the
cross section per channel is anticipated to be somewhat smaller and
therefore one needs to be able to distinguish the  
difference between that and the previous process.  In fact, a
comprehensive calculation of selectron pair-production 
at LEP 2 has been done \cite{twenty} with care taken in
delineating this boundary.
\bigskip

\begin{itemize}
\item {\bf Bottomline}
\end{itemize}
\smallskip

In a nutshell, there is no direct evidence for weak-scale
supersymmetry yet.    
Though the $R_{b,c}$ data are interesting, too
much need not be read into them.  In 
particular, any negative $\delta R_c$ has to go away first.  LEP 2
will be a crucial probe on charginos.  If no chargino is seen there,
one might forego the ``round the corner'' option.  Then a pessimist
believer might expect supersymmetry to manifest itself through the 
``decoupling'' scenario and just wait for the LHC.
\bigskip

\begin{itemize}
\item {\bf Acknowledgement} 
\end{itemize}
\smallskip

I am grateful to Gautam Bhattacharyya, Mike Bisset, Debajyoti
Choudhury, Manuel Drees, Atul Gurtu and D.P. Roy for many 
discussions and much help.  I am specially indebted to Manual Drees
for a critical 
reading of the manuscript.  I would like to thank Amitava Datta and
Partha Ghose for organizing a stimulating workshop.

\end{document}